# Magnetic relaxation of superconducting YBCO samples in weak magnetic fields

V.P. Timofeev, A.N. Omelyanchouk

*B.Verkin Institute for Low Temperature Physics & Engineering National Academy of Sciences of Ukraine,
47, Lenin Ave., 61103 Kharkov, Ukraine*
E-mail: timofeev@ilt.kharkov.ua

For the first time, magnetization of high-$T_c$ samples with different crystalline structure and its isothermal relaxation is studied at very weak constant fields ($H \leq 0.1$ Oe) for temperatures close to the critical ones. Essential influence of twin boundaries in YBCO single crystals on magnetic relaxation rate is shown. An estimation of effective pinning potential is made in the framework of the collective pinning model.

The main attention in the experimental study of magnetic flux pinning in type-II superconductors (SC) is paid to define the maximum current-carrying capacities in strong magnetic fields [1,2]. Application of highly sensitive SQUIDs in noncontact magnetic susceptibility measurements allows one to perform research in magnetic fields of small values (0.01-0.1 Oe) and even to observe spontaneous magnetic moments [3,4]. Study of the dynamics of magnetic vortices in weak magnetic fields (<< 1 Oe) is especially important for noise reduction in high-temperature superconducting (HTSC) SQUIDs. Since the self-noise of sensors is often associated with creep and hopping of vortices, its reduction is partially achieved by creation of artificial pinning centers [5].

The purpose of this paper is to present the features of magnetization $M(T, t)$ behavior of $YBa_2Cu_3O_{7-x}$ (YBCO) single crystals in very weak constant magnetic fields ($H << 0.5$ Oe) in the SC transition region. Samples of different crystalline structure are investigated to reveal the influence of thermally activated transformation of Josephson weak links on the potential of pinning centers in a system of unidirectional twin boundary planes. The main measurements are performed at temperatures near 77 K so that the obtained results could be used for the development of superconducting electronics cooled with liquid nitrogen.

The resistive state in the studied superconductors arises when magnetic vortices start moving once the acting Lorentz forces exceed the pinning ones. Due to these forces and to the effect of thermal activation occurring with probability $\sim \exp(-U/kT)$ the vortices start to move and dissipation of energy appears ($U$ is the effective activation energy of vortex hopping being equal to the average depth of pinning potential; $k$ is the Boltzmann constant; $T$ is the temperature). These processes also determine the value of the superconductor critical current ($I_c$).

An ideal superconductor placed in a weak magnetic field should be in the Meissner state. In real finite-size superconductors penetration of a magnetic field into a sample starts even at $H << H_{c1}$ ($H_{c1}$ is the first critical field of an ideal defectless superconductor of ellipsoidal shape) due to surface and volumetric defects at temperatures close to the critical one [6]. The thermally activated creep of vortices results in redistribution and damping of supercurrents, as well as in relaxation of $M$.

The data on magnetic relaxation in superconductors can be used for obtaining major parameters of the vortex pinning mechanism. Thus in the simplest case the effective depth of a pinning potential can be estimated from measurements of an isothermal relaxation $M(t)$ rate:

$$1/M_0 \, (dM/d\ln t) = - kT/U, \qquad (1)$$

where $M_0$ is the initial value of magnetization, for which one usually takes the magnetization in the Bean critical state [2]. However almost all published data on magnetic relaxation of HTSCs have been obtained in strong magnetic fields (hundreds of Oe or even several kOe), when the essential role is played by complex processes in rigid, well-formed lattice of magnetic vortices. According to the theory of collective pinning, as was demonstrated in Ref. 7, in weak magnetic fields the creep of noninteracting vortices is realized, the average velocity of flux lines does not depend on magnetic field value, and the measured data are less sensitive to the magnetic field deviation from the *c*-direction of the HTSC sample.

As the main object of study impurity-free oriented YBCO single crystals were chosen [8]. Annealing in an oxygen flow needed for obtaining optimal doping causes transformation of the initial structure of the crystals and, as a result, formation of twin boundaries. To study the influence of planar defects on pinning processes we have selected two sets of samples: one set comprises samples with unidirectional twin boundaries oriented along the c-axis throughout the crystal; the samples from the other set have domains of differently oriented twin boundaries. Dimensions of the samples are close to $1 \times 1$ mm$^2$, and their thickness is about 0,015 mm. For comparison purposes, we discuss the magnetic responses and the magnetic relaxation of polycrystalline YBCO samples with randomly oriented grains and textured $Bi_2Sr_2Ca_2Cu_3O_{10}$ ceramics.

Magnetization and magnetic relaxation in the region of the superconducting phase transition in weak magnetic fields was studied with the help of a magnetic susceptometer based on liquid-helium cooled SQUID gradiometer. The standard technique for $M(T,t)$ measurement in a homogeneous DC magnetic field of a solenoid was used. The residual magnetic field of the Earth in the working area was shielded and did not exceed 0.5 mOe. This allowed us to cool a sample and transfer it to the superconducting state in zero field (ZFC method), which is preferred for the study of $M(T)$. To analyze $M(t)$, a sample was cooled to the temperature of $\approx 77$ K in a magnetic field of

≈ 0.5 Oe (Field Cooling method), then the field was shielded, the selected $T$ was set, and the behavior of $M(t)$ was recorded.

Fig.1a shows the normalized $M(T)$ dependence for one of the studied YBCO single crystals with unidirectional twin boundaries at heating in the region of the superconducting transition. The magnetic field of the solenoid is parallel to the $c$ axis of the single crystal and has the value of 8.2 A/m (≈ 0.1 Oe). For such orientation, the field is parallel to the twin boundary planes and the pinning of Abrikosov vortices is most effective. As it can be seen from Fig. 1a, the curve of the superconducting transition, in contrast to the resistivity data from Ref. 8, is nonmonotonic and occupies a considerable temperature interval, $\Delta T \approx 5$ K. The $M(T)$ curve displays a smoothed step, which cannot be explained within the theory of vortex lattice melting [1] if one takes into consideration the low values of $H$ used in experiment. Figure 1b shows the $M(t)$ dynamics in time of this single crystal. For the sample temperature of 78 K, the magnetization $M(t)$ is characterized by a non-logarithmic dependence for low time values with subsequent passage at higher times to the behavior described by the Anderson-Kim flux-creep theory. The value of $U$ estimated by expression (1) is equal to about 0.2 eV, which is in good agreement with the data of other researchers for HTSC samples in strong magnetic fields [2]. The figure under consideration also shows the $M(t)$ behavior of the same sample at low temperatures ($T = 6.5$ K). It is clearly seen from the figure that there is no supercurrent decay and no magnetic relaxation under these conditions.

This can be explained by an exponential decay of the thermal creep of magnetic vortices and by the presence of nonsuppressed Josephson weak links in the area of twin boundaries at low temperatures and weak magnetic fields [3]. The additional argument for latter statement is the Fig. 2a with the temperature dependence of the normalized magnetization of the YBCO single crystal with blocks of multidirectional twin planes. The block boundaries with orthogonal twin planes promote formation of powerful pinning centers. The superconducting phase transition becomes sharper ($\Delta T \approx 0.3$ K) and the magnetic relaxation of this sample is not observed at 78 K.

For comparison we have studied high-temperature polycrystalline $YBa_2Cu_3O_{7-x}$ and $Bi_2Sr_2Cu_3O_{10}$ samples in weak magnetic fields. Superconducting grains of a polycrystalline sample are linked randomly by barriers of different transparency and form statistically distributed networks of circulating induced or spontaneous supercurrents. The latter can exist under certain conditions in chaotic current contours of HTSC causing paramagnetic responses. In Fig. 2b, the behavior of a normalized magnetization is shown for one of the polycrystalline $Bi_2Sr_2Ca_2Cu_3O_{10}$ samples at 77.5 K. Anisotropy of grains is responsible for the complex microstructure of polycrystalline superconductors and favors for the existence of orbital glass. Magnetic vortices can penetrate through grains and grain boundaries allowing supercurrents to flow via loops closing one or several granules.

The results of our experiments with YBCO single-crystal samples suggest that twinning planes create conditions for the formation of similar Josephson networks with randomly distributed parameters. The twin boundaries include CuO layers containing oxygen vacancies, have strong local suppression of the superconducting order parameter and produce system of field sensitive pinning centers.

Thus the temperature dependence of magnetization and magnetic relaxation of different high-temperature samples is investigated in very weak magnetic fields ($H \leq 0.1$ Oe) at temperatures close to the critical ones. A profound effect of twin boundaries in single crystals on magnetization is shown. An effective pinning potential of optimally doped YBCO single crystals is estimated within the model of collective pinning.

The authors are grateful to A.V. Bondarenko for the grown single crystals and to V.N. Samovarov for helpful discussions.

**References**

1. G.Blatter, M.V.Feigel'man, V.B.Geshkenbein, A.I.Larkin, and V.M.Vinokur, *Rev. Mod. Phys.* **66**, 1125 (1994).
2. Y.Yeshurun, A.P.Malozemoff, and A.Shaulov, *Rev. Mod. Phys.* **68**, 911 (1996).
3. V.P.Timofeev, A.V.Bondarenko, *Fiz. Nizk. Temp.* **30**, 810 (2004) [*Low Temp. Phys.* **30**, 610 (2004)].
4. Yu.A.Kolesnichnko, A.N.Omel'yanchouk, and A.M.Zagoskin, *Fiz. Nizk. Temp* 30, 714 (2004) [*Low Temp. Phys.*, **30**, 535 (2004)].
5. P.Selders, A.M.Castellanos, M.Vanpel, and R.Wordenweber, *IEEE Trans. on Appl. Superconductivity* **9**, 2967 (1999).
6. R.Liang, D.A.Bonn, W.N.Hardly, and D.Broun, *Phys. Rev. Lett.* **94**, 117001 (2005).
7. A.V.Bondarenko, A.A.Prodan, M.A.Obolenskii, R.V.Vovk, and T.R.Arouri, *Fiz. Nizk. Temp.* **27**, 463 (2001) [*Low Temp. Phys.* **27**, 339 (2001)].
8. M.A.Obolenskii, A.V.Bondarenko, and M.O.Zubareva, *Fiz. Nizk. Temp.* **15**, 1152 (1989) [*Low Temp. Phys.* **15**, 635 (1989)].

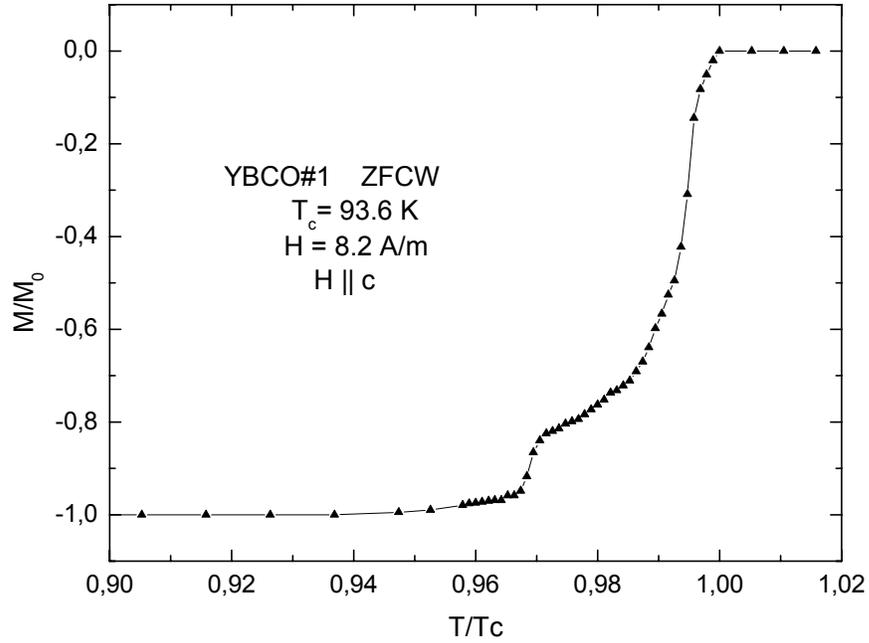

(a)

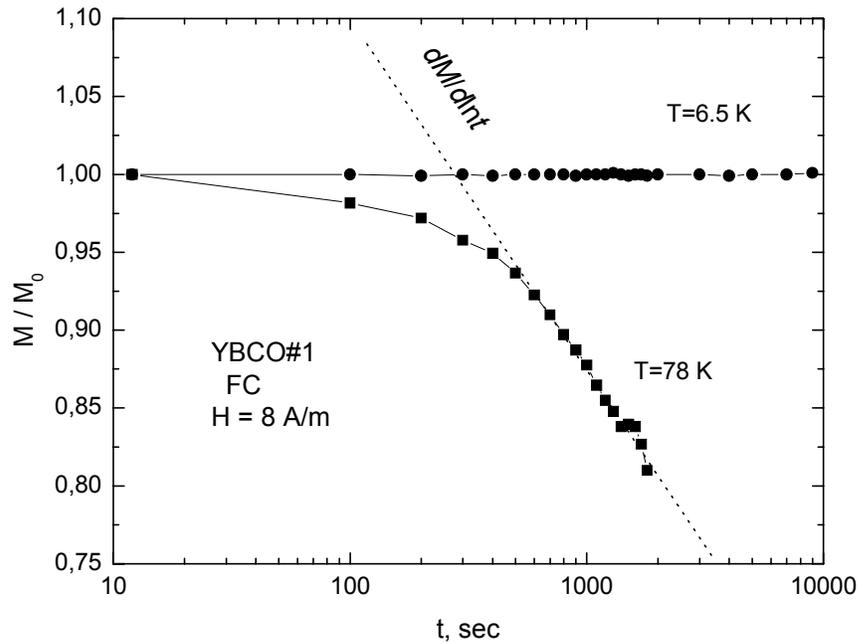

(b)

Fig.1. Normalized temperature dependence of magnetization $M(T)$ for YBCO single crystal #1 with unidirectional twin planes in the region of the superconducting phase transition (a); magnetization dynamics $M(t)$ of the same sample in the magnetic field of 8 A/m ($\approx$ 0.1 Oe) at two different temperatures (b). The linear part of the dependence used for the estimation of effective pinning potential is shown as a dotted line. ZFCW – zero field cooled warming.

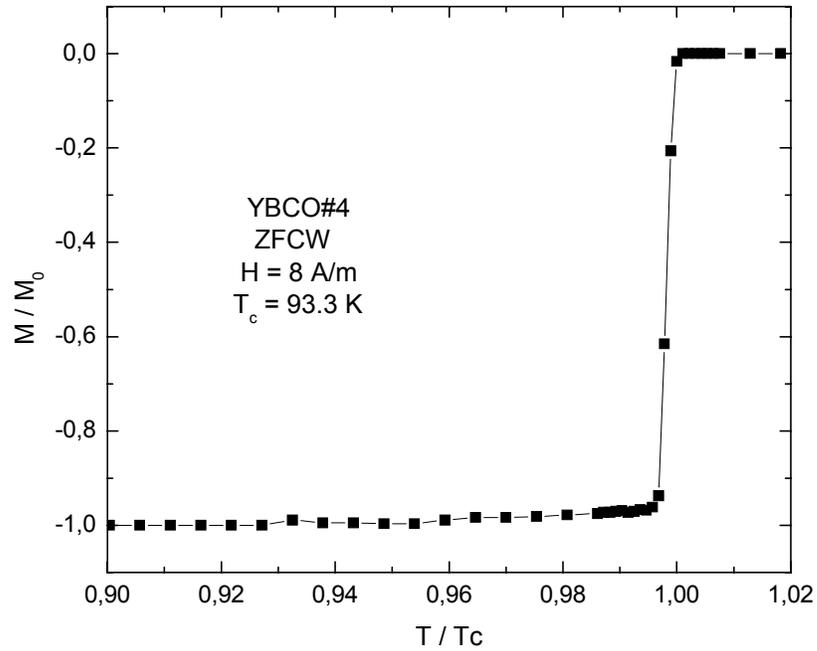

(a)

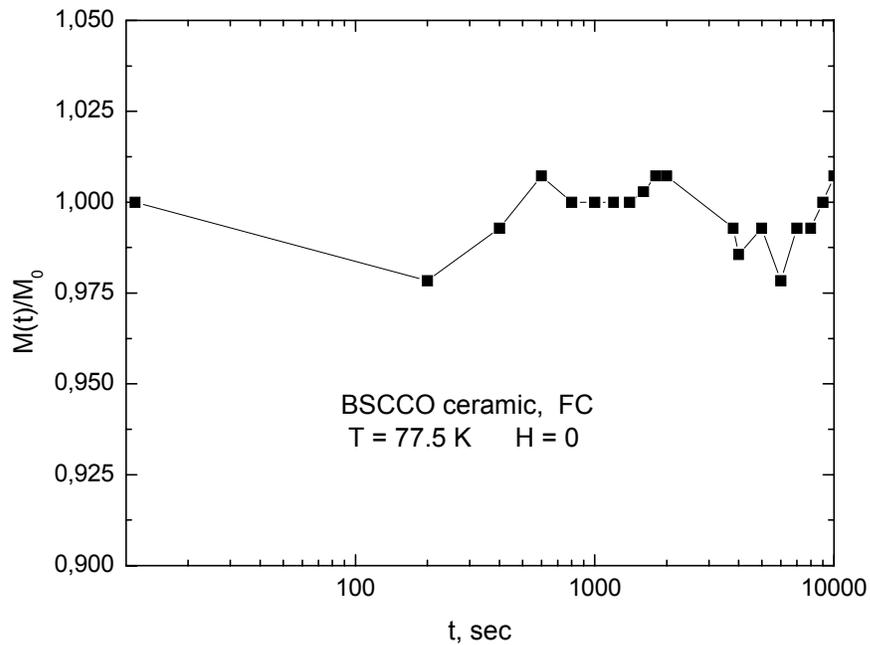

(b)

Fig.2. Normalized temperature dependence of magnetization $M(T)$ for YBCO single crystal #4 with multidirectional twin planes at the region of the superconducting phase transition in the magnetic field of 8 A/m ($\approx$ 0.1 Oe) which is parallel to the *c*-axis of the crystal (a); time dependence of magnetization of the polycrystalline $Bi_2Sr_2Ca_2Cu_3O_{10}$ sample in zero magnetic field at 77.5 K(b).